\documentclass[10pt,conference]{IEEEtran}
\IEEEoverridecommandlockouts
\usepackage{cite}
\usepackage{amsmath,amssymb,amsfonts}
\usepackage{algorithmic}
\usepackage{graphicx}
\usepackage{textcomp}
\usepackage{xcolor}
\usepackage{booktabs}
\usepackage{multirow}
\usepackage{eso-pic}
\usepackage{eso-pic,rotating,xcolor}
\usepackage{fancyhdr}

\usepackage[draft]{pdfcomment} % https://ctan.org/pkg/pdfcomment?lang=en 
\usepackage[hyperfirst=false,acronym,sort=none,shortcuts,nopostdot,style=super,nonumberlist,toc,nogroupskip]{glossaries}
\usepackage{xcolor} % used for colored acronyms

\usepackage{xspace}

\glsdisablehyper % to disable hyperlinks on all acronyms

\definecolor{Sepia}{RGB}{112,66,20}
\definecolor{AcColor}{rgb}{0,0,0}

\newcommand{\accolor}[1]{\textcolor{AcColor}{#1}} % subsequent uses
\newacronymstyle{myacro}
{%
  \GlsUseAcrEntryDispStyle{long-short}%
}%
{%
  \GlsUseAcrStyleDefs{long-short}%
}

\setacronymstyle{myacro}

\let\xx\tip
\let\xxx\tips
\newacronym{adc}{ADC}{Analog to Digital Converter}
\newacronym{admm}{ADMM}{Alternating Direction Method of Multipliers}
\newacronym{asic}{ASIC}{Application Specific Integrated Circuit}
\newacronym{cem}{CEM}{Cross-Entropy Method}
\newacronym{cots}{COTS}{Commodity Off-The-Shelf}
\newacronym{cpu}{CPU}{Central Processing Unit}
\newacronym{cnn}{CNN}{Convolutional Neural Network}
\newacronym{ddp}{DDP}{Differential Dynamic Programming}
\newacronym{dnn}{DNN}{Deep Neural Network}
\newacronym{dns}{DNS}{Deep Noise Suppression}
\newacronym{dof}{DOF}{Degree Of Freedom}
\newacronym{dram}{DRAM}{Dynamic RAM}
\newacronym{fc}{FC}{Fully Connected}
\newacronym{fpga}{FPGA}{Field Programmable Gate Array}
\newacronym{gpu}{GPU}{Graphics Processing Unit}
\newacronym{gru}{GRU}{Gated Recurrent Unit}
\newacronym{hil}{HIL}{Hardware In the Loop}
\newacronym{ip}{IP}{Interior Point}
\newacronym{ipb}{IP block}{Intellectual Property Block}
\newacronym{lstm}{LSTM}{Long Short Term Memory}
\newacronym{ltv}{LTV}{Linear Time Varying}
\newacronym{mac}{MAC}{Multiply-Accumulate}
\newacronym{mae}{MAE}{Median Angular Error}
\newacronym{mlp}{MLP}{Multilayer Perceptron}
\newacronym{mee}{MEE}{Median Endpoint Error}
\newacronym{mpc}{MPC}{Model Predictive Control}
\newacronym{mpcer}{MPC}{Model Predictive Controller}
\newacronym{mppi}{MPPI}{Model Predictive Path Integral}
\newacronym{nc}{NC}{Neural Controller}
\newacronym{nmpc}{NMPC}{Nonlinear Model Predictive Control}
\newacronym{nn}{NN}{Neural Network}
\newacronym{npc}{NPC}{Neural Predictive Control}
\newacronym{npu}{NPU}{Neural Processing Unit}
\newacronym{ode}{ODE}{Ordinary Differential Equation}
\newacronym{pcb}{PCB}{Printed Circuit Board}
\newacronym{pd}{PD}{Proportional Derivative}
\newacronym{pid}{PID}{Proportional Integral Derivative}
\newacronym{pl}{PL}{Programmable Logic}
\newacronym{ps}{PS}{Processing System}
\newacronym{pso}{PSO}{Particle Swarm Optimization}
\newacronym{pwm}{PWM}{Pulse Width Modulation}
\newacronym{qp}{QP}{Quadratic Programming}
\newacronym{rl}{RL}{Reinforcement Learning}
\newacronym{rnn}{RNN}{Recurrent Neural Network}
\newacronym{ros}{ROS}{Robot Operating System}
\newacronym{rpgd}{RPGD}{Resampling Parallel Gradient Descent}
\newacronym{slp}{SLP}{Single Layer Perceptron}
\newacronym{sm}{SM}{Supplementary Material}
\newacronym{snr}{SNR}{Signal-to-Noise Ratio}

\newacronym[description={System on Chip; FPGA with embedded programmable processor}]{soc}{SoC}{System on Chip}
\newacronym{sqp}{SQP}{Sequential Quadratic Programming}
\newacronym{sram}{SRAM}{Static RAM}
\newacronym{usb}{USB}{Universal Serial Bus}
\newacronym{vga}{VGA}{Video Graphics Adaptor}
\newacronym{uart}{UART}{Universal Asynchronous Receiver/Transmitter}
\newacronym{xla}{XLA}{Accelerated Linear Algebra}

\newacronym{ssm}{SSM}{State Space Model}
\newacronym{film}{FiLM}{Feature-wise Linear Modulation}
\newacronym{dprnn}{DPRNN}{Dual-Path RNN}
\newacronym{sota}{SOTA}{State Of The Art}
\newacronym{ec}{EC}{Embedding Concatenation}
\newacronym{ssmm}{SSMM}{SSM Modulation}
\newacronym{sisnr}{SISNR}{Scale-Invariant Signal-to-Noise Ratio}
\newacronym{lr}{lr}{learning rate}
\newacronym{se}{SE}{Speech Enhancement}
\newacronym{ola}{OLA}{Overlap-and-Add}
\newacronym{anc}{ANC}{Active Noise Control}

\def\method{{\bf Slow{\it Fast}} } %AutoFit

\def\BibTeX{{\rm B\kern-.05em{\sc i\kern-.025em b}\kern-.08em
    T\kern-.1667em\lower.7ex\hbox{E}\kern-.125emX}}

\newcommand{\newlineauthors}{%
\end{@IEEEauthorhalign}\hfill\mbox{}\par
\mbox{}\begin{@IEEEauthorhalign}}

\begin{document}

\title{{\huge Modulating State Space Model with {\bf Slow{\it Fast}} Framework for Compute-Efficient Ultra Low-Latency Speech Enhancement}

\thanks{This work was supported by a research contract from Meta Reality Labs Research.} 
}

\author{
 \IEEEauthorblockN{Longbiao Cheng$^1$, Ashutosh Pandey$^2$, Buye Xu$^2$, Tobi Delbruck$^1$, Vamsi Krishna Ithapu$^2$, Shih-Chii Liu$^1$}\\
 \IEEEauthorblockA{
$^1$ Institute of Neuroinformatics, University of Zurich and ETH Zurich, Zurich, Switzerland\\
$^2$ Reality Labs Research, Meta, Redmond, United States\\
\{longbiao, tobi, shih\}@ini.uzh.ch, \{apandey620, xub, ithapu\}@meta.com
}
}

\maketitle

\begin{abstract}
Deep learning-based speech enhancement (SE) methods often face significant computational challenges when needing to meet low-latency requirements because of the increased number of frames to be processed. This paper introduces the {\bf Slow{\it Fast}} framework which aims to reduce computation costs specifically when low-latency enhancement is needed. The framework consists of a slow branch that analyzes the acoustic environment at a low frame rate, and a fast branch that performs SE in the time domain at the needed higher frame rate to match the required latency. Specifically, the fast branch employs a state space model where its state transition process is dynamically modulated by the slow branch. Experiments on a SE task with a 2 ms algorithmic latency requirement using the Voice Bank + Demand dataset show that our approach reduces computation cost by 70\% compared to a baseline single-branch network with equivalent parameters, without compromising enhancement performance. Furthermore, by leveraging the {\bf Slow{\it Fast}} framework, we implemented a network that achieves an algorithmic latency of just 62.5 $\mu$s (one sample point at 16 kHz sample rate) with a computation cost of 100 M MACs/s, while scoring a PESQ-NB of 3.12 and SISNR of 16.62.
\end{abstract}


\section{Introduction\label{sec:intro}}
Low-latency speech signal processing is crucial for devices and applications such as hearing aids, communication devices, and immersive virtual reality, where minimizing delays in the output of processed speech signal is essential to ensure optimal user experience~\cite{stone2008tolerable, gupta2022augmented}. Recent deep learning-based \xx{se} algorithms have focused extensively on solutions to meet these low-latency requirements. For instance, causal networks have been utilized to reduce the number of lookahead samples needed~\cite{nicolson2020masked, pandey2021dense, fan2022real, oostermeijer2021lightweight, takeuchi2020real, cheng2024dynamic}. Additionally, various methods have been developed to reduce the algorithmic latency, achieving as low as 2 ms. Several studies~\cite{luo2018tasnet, luo2020dual, tu2021two,gajecki2021binaural} employ symmetric analysis-synthesis windows, where both the analytical window used to frame the input signal and the synthesis window used in the \xx{ola} process have the same length to match the algorithmic latency requirements. In contrast, other studies~\cite{wang2021deep, wang2022stft, pandey2023simple} propose the use of asymmetric windows~\cite{mauler2007low, wood2019unsupervised}, where a larger window size is used for analysis to capture more extensive acoustic information, while a smaller window is used during synthesis to reduce algorithmic latency.

As latency requirements tighten, more frames must be processed per unit of time due to the shorter hop size needed for framing inputs. For example, reducing algorithmic latency from 20 ms to 2 ms requires processing 10$\times$ more frames, which causes a tenfold increase in computation cost for existing methods. This poses significant challenges for deploying current low algorithmic latency networks on edge devices. Meanwhile, reducing the hop size also leads to a decreased differences between consecutive frames, thus increasing information redundancy between adjacent inputs. However, despite this redundancy, current methods still run the same network for every input frame, resulting in a substantial amount of redundant computations.

\begin{figure*}[th]
  \centering
  \includegraphics[width=\linewidth]{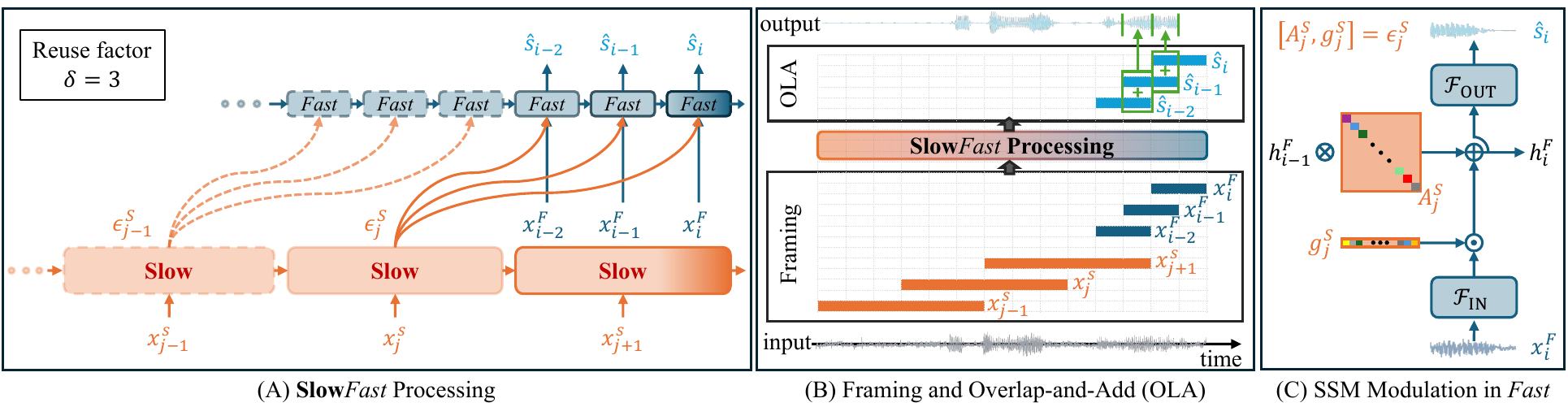}
  \vspace*{-7.5mm}
  \caption{
   Illustration of proposed \method framework for compute-efficient low-latency speech enhancement.  
   (A) \method Processing when $\delta=3$.  : The slow branch (orange, bottom) operates at a lower frame rate, the fast branch (blue, top) operates at a higher rate. (B) Framing and OLA Process: The slow branch processes longer segments with a larger hop size, while the fast branch processes shorter segments with a smaller hop size. The enhanced speech is obtained by doing the \xx{ola} on the fast branch outputs. (C) SSM Modulation: The slow branch modulates the state transition process in the fast branch. 
  }
  \label{fig:slowfast}
% \vspace{-2.5mm}
\end{figure*}

In this paper, we propose the \method framework designed to reduce the computation cost of \xx{se} networks under low-latency requirement while maintaining enhancement performance. 
To the best of our knowledge, this is the first framework {\bf enabling single sample-level algorithmic latency speech enhancement} at an extremely low computation cost.
The key contributions of this work are as follows:
\begin{itemize}
\item 
We propose a dual-branch \xx{se} framework in which the slow branch, being relatively computation-heavy, operates at a lower frame rate, while only the fast branch runs at a higher rate, thereby reducing overall computational costs.
\item We introduce a lightweight \xx{ssm}-based fast branch. 
In the \xx{ssm}, the state transitions are dynamically modulated according to the speech and noise characteristics modeled by the slow branch. 
\item Evaluation results on a \xx{se} task demonstrate that networks designed with the \method framework can achieve single sample-level algorithmic latency with a computation cost of approximately 100 M \xxx{mac} per second.
Furthermore, for a 2 ms algorithmic latency requirement scenario, evaluations on the Voice Bank + DEMAND dataset show a 70\% reduction in network computation cost, while preserving the quality and intelligibility of the enhanced speech. 
\end{itemize}

\section{Proposed Method}

This section outlines the processing pipeline of the \method framework and introduces the proposed \xx{ssmm}-based slow-fast integration method.

\subsection{\method Framework}
As depicted in Fig.~\ref{fig:slowfast}, the \method framework employs a dual-branch architecture with distinct window lengths and hop sizes.

The input frames to the fast and slow branches are denoted as ${\bf x}^F = [x^F_0, \dots, x^F_i]$ and ${\bf x}^S = [x^S_0, \dots, x^S_j]$, respectively. Here, $i \in [0, N_F)$ and $j \in [0, N_S)$ represent the frame indices for the fast and slow branches, where $N_F$ and $N_S$ are the total number of input frames for each branch.
%Under this setup, 
The workflow of the \method framework can be summarized as follows:
\begin{align}
{\boldsymbol \epsilon}^S &= f_{\rm \bf Slow}({\bf x}^S), \\
\hat{\bf s} \ \  &= f_{ Fast}({\bf x}^F; {\boldsymbol \epsilon}^S).
\end{align}
Here, $f_{\rm \bf Slow}(\cdot)$ represents the processing function of the slow branch, producing ${\boldsymbol \epsilon}^S = [\epsilon^S_0, \dots, \epsilon^S_j]$. $f_{Fast}(\cdot)$ represents the processing function of the fast branch, enhancing each $x^F_i$ with utilizing slow branch output and generating a corresponding enhanced speech frame $\hat{s}_i$, resulting in $\hat{\bf s} = [\hat{s}_0, \dots, \hat{s}_i]$. The final enhanced speech is obtained by overlapping and adding the frames from $\hat{\bf s}$.

The algorithmic latency of proposed \method is determined by the length of $\hat{s}_i$. To meet the latency requirement and in parallel, optimizing the computation efficiency, each $x^F_i$ to the fast branch is framed with a window length $L_F$, which is aligned with the latency requirement, and a hop size $\Delta_F$.
Meanwhile, the slow branch frames $x^S_j$ with a larger hop size $\Delta_S$, where $\Delta_S > \Delta_F$. This configuration reduces the number of frames that the slow branch needs to process compared to the fast branch, thereby lowering the computation cost relative to conventional single branch \xx{se} frameworks. Additionally, the slow branch employs a larger window length $L_S$ ($L_S > L_F$), enabling it to capture more comprehensive characteristics of the acoustic environment, including both noise and speech.

It is important to note that because $\Delta_F < \Delta_S$, $N_F$ is approximately $\delta = \Delta_S // \Delta_F$ times larger than $N_S$. This means that for every $\delta$ consecutive input frames in the fast branch, the same $\epsilon^S_j$ will be reused, thus here defining $\delta$ as the {\bf reuse factor}. For simplicity, we always set $\Delta_S = \delta \cdot \Delta_F$.
Moreover, as illustrated in Fig.~\ref{fig:slowfast}, $\epsilon^S_j$ should be generated without using the current inputs to the fast branch to maintain the causality. An additional benefit of this setup is that the slow branch has $\delta$ times more processing time than the fast branch, provides a possibility for deploying the slow branch on an remote device~\cite{srinivas2024knowledge}. Under all these setup, we have $j = i // \delta - 1$.

In practice, the $f_{\rm \bf Slow} (\cdot)$ can be implemented using any \xx{se} network as the backbone, followed by a \xx{fc} layer for generating $\epsilon^S_j$. The details of the fast branch are introduced in the next section.

\subsection{SSM Modulation (SSMM) For Slow-Fast Integration\label{sec:slow-fast}}
The success of \method relies on a lightweight fast branch that effectively utilizes the acoustic environment characteristics sensed by the slow branch. We achieve this through the proposed \xx{ssmm} based slow-fast integration. 

As shown in Fig.~\ref{fig:slowfast}(C), the fast branch is conceptualized as a \xx{ssm}, with the slow branch modulating its state transitions. This process can be formalized as follows:
\begin{align}
    h^F_i = A^S_j \times h^F_{i-1} + g^S_j \cdot {\mathcal{F}}_{\rm IN}\left(x^F_i \right)  {\bf {\scriptstyle(State \   Equation)}}  &\\
    \hat{s}_i = {\mathcal{F}}_{\rm OUT}\left(h^F_i \right) \ \ \   \quad \quad \quad \quad \quad  {\bf {\scriptstyle(Output \ Equation)}} &
\end{align}
where $[A^S_j, g^S_j] = \epsilon^S_j$ are generated by the slow branch,  $h^F_i \in \mathbb{R}^H$ is the hidden state with a dimension of $H$.
${\mathcal{F}}_{\rm IN}(\cdot)$ maps the time domain noisy input $x^F_i$ to the hidden space. The vector $g^S_j \in \mathbb{R}^H $ gates the noisy input based on the noise characteristics modeled by the slow branch, performing pre-enhancement in the hidden domain. $A^S_j \in \mathbb{R}^{H\times H}$, represents the state transition process informed by the speech characteristics sensed by the slow branch, capturing the evolution dynamics of the speech signal from the previous frame $i-1$ to the current frame $i$. ${\mathcal{F}}_{\rm OUT}(\cdot)$ then maps the speech representation $h^F_i$ back to time domain as the enhanced speech frame $\hat{s}_i$.

Inspired by recent advances in structured \xxx{ssm}~\cite{gu2021efficiently, orvieto2023resurrecting}, we simplify the state transition matrix $A^S_j$ to a diagonal matrix. This simplification reduces the parameter space, with $\epsilon^S_j$ containing only $2H$ elements. As a result, $\epsilon^S_j$ can be generated without imposing significant computation overhead on the slow branch.
To maintain computation efficiency, %unless otherwise specified, 
both ${\mathcal{F}}_{\rm IN}(\cdot)$ and ${\mathcal{F}}_{\rm OUT}(\cdot)$ are implemented as a single \xx{fc} layer.

\begin{figure}[t]
  \centering
  \includegraphics[width=0.95\linewidth]{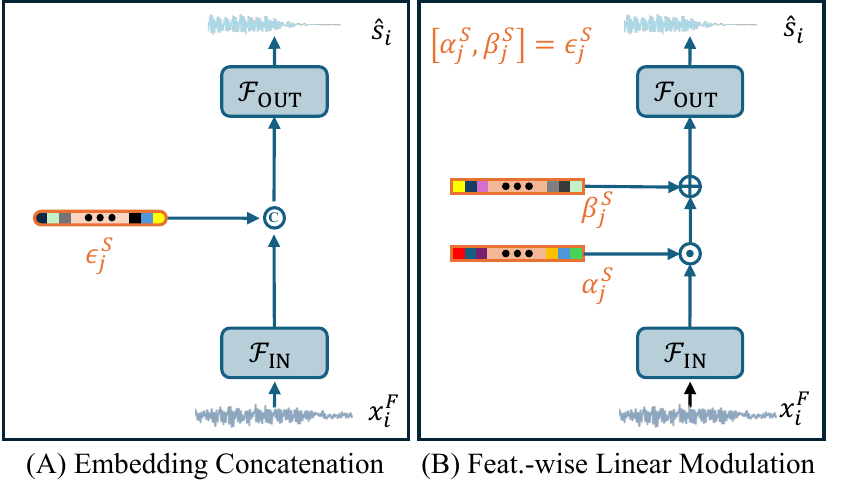}
  \vspace*{-4mm}
  \caption{
  %Two other methods for integrating the {\bf Slow} and {\it Fast} branches.
  Two other methods investigated in this work for integrating the {\bf Slow} and {\it Fast} branches.
  (A) Embedding Concatenation: The output of the slow branch is used as an additional feature in fast branch. (B) Feature-wise Linear Modulation: The slow branch generates two vectors that scale and shift the fast branch features. 
  %Throughout the figure, orange elements represent parameters or features generated by slow branch, while blue ones denote operations in fast branch.
  }

  \label{fig:conditioning}
% \vspace{-2.5mm}
\end{figure}

As shown in Fig.~\ref{fig:conditioning}, in addition to the proposed SSMM, we investigated two other slow-fast integration methods: 1) \xx{ec} and 2) \xx{film}. The \xx{ec} method directly concatenates $\epsilon^S_j \in \mathbb{R}^H$ with the fast branch features for final enhancement. \xx{film} applies a linear transformation~\cite{perez2018film}, including scaling and shifting using $\alpha^S_j\in \mathbb{R}^H$ and $\beta^S_j\in \mathbb{R}^H$, to the fast branch features, with parameters generated by the slow branch. Therefore, in \xx{ec} and \xx{film} methods, $\epsilon^S_j$ contains $H$ and $2H$ elements, respectively.

\subsection{Loss Function}
The loss function used in training all networks is:
\begin{equation}
    \mathcal{L} = \lambda_{1}\mathcal{L}_{\rm SpecMSE} + \lambda_2\mathcal{L}_{\rm SISNR} + \lambda_3\mathcal{L}_{\rm PESQ} + \lambda_4\mathcal{L}_{\rm STOI}
    \label{eq:loss}
\end{equation}
where $\mathcal{L}_{\rm SpecMSE}$ represents the mean squared error on the magnitude, real, and imaginary parts of the spectrum between the enhanced speech and the target~\cite{cheng2021fscnet}. $\mathcal{L}_{\rm SISNR}$ is the negative \xx{sisnr} between the enhanced and target speech. $\mathcal{L}_{\rm PESQ}$ and $\mathcal{L}_{\rm STOI}$ are perceptual-based loss functions proposed in \cite{martin2018deep} and \cite{zhang2018training}, respectively. The scaling factors $\lambda_{1}, \lambda_{2}, \lambda_{3}$, and $\lambda_{4}$ are used to balance the influence of each loss component.

% Please add the following required packages to your document preamble:
% \usepackage{booktabs}
\begin{table*}[t]
\centering
\caption{Results (Mean {\footnotesize ± Standard Deviation}) of Six SE Measures Under 2 ms Algorithmic Latency Requirement}
\vspace*{-2.5mm}
\begin{tabular}{ clcccccccc }
\toprule
ID  & \multicolumn{1}{c}{Method} & $\delta$ & \xxx{mac} (M/s) & PESQ-NB              & PESQ-WB              & STOI (\%)              & ESTOI (\%)             & HASPI (\%)             & SISNR                 \\ \midrule \midrule
 & noisy          & -        & -    & 2.87                 & 1.97                 & 92.10                   & 78.66                 & 97.12                 & 8.45                  \\ \midrule
0   & Single-Branch                  & -        & 126  & 3.25 \footnotesize{± 0.01}          & 2.59 \footnotesize{± 0.03}          & 93.14 \footnotesize{± 0.06}          & 82.56 \footnotesize{± 0.10}          & 98.85 \footnotesize{± 0.12}          & 17.79 \footnotesize{± 0.12}          \\ \midrule
1.A & SlowFast-EC                   & 1        & 110  & 3.24 \footnotesize{± 0.02}          & 2.45 \footnotesize{± 0.03}          & 93.24 \footnotesize{± 0.06}          & 82.47 \footnotesize{± 0.14}          & 98.47 \footnotesize{± 0.11}          & 17.72 \footnotesize{± 0.03}          \\ 
1.B & SlowFast-FiLM                 & 1        & 110  & 3.27 \footnotesize{± 0.01}          & 2.61 \footnotesize{± 0.01}          & 93.37 \footnotesize{± 0.06}          & 83.29 \footnotesize{± 0.15}          & 99.06 \footnotesize{± 0.05}          & 18.11 \footnotesize{± 0.03}          \\
1.C & SlowFast-SSMM                 & 1        & 110  & \textbf{3.30 \footnotesize{± 0.02}} & \textbf{2.66 \footnotesize{± 0.02}} & \textbf{93.38 \footnotesize{± 0.08}} & \textbf{83.42 \footnotesize{± 0.18}} & \textbf{99.10 \footnotesize{± 0.07}} & \textbf{18.15 \footnotesize{± 0.02}} \\ \midrule
2.A & SlowFast-EC                   & 2        & 57   & 2.90 \footnotesize{± 0.01}          & 2.07 \footnotesize{± 0.00}          & 92.70 \footnotesize{± 0.05}          & 80.93 \footnotesize{± 0.07}          & 97.25 \footnotesize{± 0.08}          & 16.00 \footnotesize{± 0.04}          \\
2.B & SlowFast-FiLM                 & 2        & 57   & 3.19 \footnotesize{± 0.03}          & 2.50 \footnotesize{± 0.03}          & 93.12 \footnotesize{± 0.05}          & 82.66 \footnotesize{± 0.10}          & 98.50 \footnotesize{± 0.11}          & 17.80 \footnotesize{± 0.07}          \\
2.C & SlowFast-SSMM                 & 2        & 57   & \textbf{3.26 \footnotesize{± 0.02}} & \textbf{2.61 \footnotesize{± 0.02}} & \textbf{93.18 \footnotesize{± 0.05}} & \textbf{82.89 \footnotesize{± 0.06}} & \textbf{98.96 \footnotesize{± 0.07}} & \textbf{17.89 \footnotesize{± 0.09}} \\ \midrule
3.A & SlowFast-EC                   & 3        & 39   & 2.77 \footnotesize{± 0.01}          & 2.07 \footnotesize{± 0.01}          & 92.01 \footnotesize{± 0.02}          & 79.39 \footnotesize{± 0.06}          & 96.74 \footnotesize{± 0.09}          & 15.69 \footnotesize{± 0.08}          \\
3.B & SlowFast-FiLM                 & 3        & 39   & 3.15 \footnotesize{± 0.02}          & 2.45 \footnotesize{± 0.02}          & 92.92 \footnotesize{± 0.03}          & 82.20 \footnotesize{± 0.07}          & 98.46 \footnotesize{± 0.09}          & 17.64 \footnotesize{± 0.02}          \\
3.C & SlowFast-SSMM                 & 3        & 39   & \textbf{3.24 \footnotesize{± 0.01}} & \textbf{2.59 \footnotesize{± 0.02}} & \textbf{93.10 \footnotesize{± 0.07}} & \textbf{82.78 \footnotesize{± 0.12}} & \textbf{98.86 \footnotesize{± 0.03}} & \textbf{17.86 \footnotesize{± 0.05}} \\ \midrule
4.A & SlowFast-EC                   & 4        & 31   & 2.73 \footnotesize{± 0.01}          & 2.06 \footnotesize{± 0.01}          & 91.60 \footnotesize{± 0.04}          & 78.43 \footnotesize{± 0.06}          & 97.28 \footnotesize{± 0.07}          & 15.22 \footnotesize{± 0.14}          \\
4.B & SlowFast-FiLM                 & 4        & 31   & 3.15 \footnotesize{± 0.03}          & 2.46 \footnotesize{± 0.02}          & 92.73 \footnotesize{± 0.06}          & 81.93 \footnotesize{± 0.13}          & 98.46 \footnotesize{± 0.12}          & 17.57 \footnotesize{± 0.03}          \\
4.C & SlowFast-SSMM                 & 4        & 31   & \textbf{3.19 \footnotesize{± 0.01}} & \textbf{2.53 \footnotesize{± 0.01}} & \textbf{92.90 \footnotesize{± 0.07}} & \textbf{82.38 \footnotesize{± 0.14}} & \textbf{98.76 \footnotesize{± 0.05}} & \textbf{17.74 \footnotesize{± 0.04}} \\ \midrule
5.A & SlowFast-EC                   & 5        & 26   & 2.71 \footnotesize{± 0.01}          & 2.02 \footnotesize{± 0.01}          & 91.32 \footnotesize{± 0.02}          & 77.70 \footnotesize{± 0.06}          & 97.37 \footnotesize{± 0.06}          & 14.74 \footnotesize{± 0.13}          \\
5.B & SlowFast-FiLM                 & 5        & 25   & 3.13 \footnotesize{± 0.02}          & 2.41 \footnotesize{± 0.02}          & 92.60 \footnotesize{± 0.05}          & 81.58 \footnotesize{± 0.11}          & 98.41 \footnotesize{± 0.03}          & 17.43 \footnotesize{± 0.04}          \\
5.C & SlowFast-SSMM                 & 5        & 25   & \textbf{3.18 \footnotesize{± 0.01}} & \textbf{2.52 \footnotesize{± 0.02}} & \textbf{92.75 \footnotesize{± 0.05}} & \textbf{82.06 \footnotesize{± 0.14}} & \textbf{98.63 \footnotesize{± 0.02}} & \textbf{17.66 \footnotesize{± 0.03}} \\ \midrule
6.A & SlowFast-EC                   & 10       & 15   & 2.71 \footnotesize{± 0.01}          & 1.88 \footnotesize{± 0.01}          & 91.12 \footnotesize{± 0.04}          & 77.23 \footnotesize{± 0.06}          & 96.77 \footnotesize{± 0.03}          & 13.36 \footnotesize{± 0.12}          \\
6.B & SlowFast-FiLM                 & 10       & 15   & 3.04 \footnotesize{± 0.01}          & 2.24 \footnotesize{± 0.01}          & 92.05 \footnotesize{± 0.08}          & 80.39 \footnotesize{± 0.14}          & 98.15 \footnotesize{± 0.07}          & 16.42 \footnotesize{± 0.04}          \\
6.C & SlowFast-SSMM                 & 10       & 15   & \textbf{3.10 \footnotesize{± 0.01}} & \textbf{2.39 \footnotesize{± 0.01}} & \textbf{92.36 \footnotesize{± 0.09}} & \textbf{81.14 \footnotesize{± 0.16}} & \textbf{98.62 \footnotesize{± 0.12}} & \textbf{17.08 \footnotesize{± 0.03}} \\ \midrule
7   & Single-Branch                 & -        & 45   & 3.09 \footnotesize{± 0.02}          & 2.39 \footnotesize{± 0.03}          & 92.72 \footnotesize{± 0.09}          & 81.37 \footnotesize{± 0.28}          & 97.97 \footnotesize{± 0.08}          & 16.91 \footnotesize{± 0.15}          \\ \midrule \midrule
8   & RNNoise~\cite{valin2018hybrid}*                 & -        & 40  & -          & 2.43          & -          & -          & -         & -              \\
9   & DeepFiltering~\cite{schroter2022low}$^\dagger$  & -        & 233    & -          & 2.65          & 93.8       & -          & -         & 14.04          \\
10   & $\rm ULCNet_{Freq}$~\cite{shetu2024ultra}*     & -        & 98  & -          & 2.54          & -          & -          & -         & 17.20          \\
\bottomrule
\multicolumn{10}{l}{*: 20 ms algorithmic latency. Adapting the network with conventional single-branch framework to meet a 2 ms latency will increase \xxx{mac} by 10x.} \\
\multicolumn{10}{l}{$^\dagger$: 8 ms algorithmic latency. Adapting the network with conventional single-branch framework to meet a 2 ms latency will increase \xxx{mac} by 4x.}
\end{tabular}
\label{tab:2ms}
% \vspace{-2mm}
\end{table*}

% Please add the following required packages to your document preamble:
% \usepackage{booktabs}
% \usepackage[table,xcdraw]{xcolor}
% Beamer presentation requires \usepackage{colortbl} instead of \usepackage[table,xcdraw]{xcolor}
\begin{table*}[t]
\centering
\caption{Results (Mean ± Standard Deviation) of \method Framework Under Single Sample-Level Algorithmic Latency Constraint}
\vspace*{-2.5mm}
\begin{tabular}{ccccccccc}
\toprule
Method                             & $\delta$ & \xxx{mac} (M/s) & PESQ-NB     & PESQ-WB     & STOI (\%)     & ESTOI (\%)    & HASPI (\%)    & SISNR        \\ \midrule \midrule
noisy                                 & -        & -    & 2.87                 & 1.97                 & 92.10                   & 78.66                 & 97.12                 & 8.45                  \\ 
SlowFast-SSMM & 16       & 105  & 3.12 ± 0.02 & 2.41 ± 0.03 & 92.57 ± 0.05 & 81.32 ± 0.17 & 97.80 ± 0.04 & 16.62 ± 0.08 \\ \bottomrule
\end{tabular}
\label{tab:sample-level}

% \vspace{-2.5mm}
\end{table*}

\section{Experimental Setup}

Experiments are conducted to demonstrate the  effectiveness of \method in reducing computation costs under a 2ms algorithmic latency requirement for \xx{se} and to determine its ability for achieving sample-level latency enhancement. 

% \subsection{Dataset}
The proposed framework is evaluated on the widely used Voice Bank + DEMAND (VBD) dataset~\cite{valentini2017noisy}. 
In VBD, samples from 28 out of 30 speakers from the Voice Bank corpus~\cite{veaux2013voice} are used for training, while the remaining 2 speakers are used for testing. 
For the training set, 10 types of noise from the DEMAND corpus~\cite{thiemann2013diverse} are mixed with clean speech for four  \xx{snr} values (0, 5, 10, and 15 dB). The test set consists of clean speech mixed with 5 types of noise, not used in the training set, and at four different \xx{snr} values (2.5, 7.5, 12.5, and 17.5 dB). All data are sampled at 16 kHz.

% \subsection{Networks}
For the 2 ms latency experiments, the parameters are set as follows: $L_F = 32$, $\Delta_F = 16$, and $\delta \in {1, 2, 3, 4, 5, 10}$. Additionally, $L_S = 2\Delta_S$ and $H = 32$. In the single sample-level latency experiment, we use $L_F = 1$ and $\Delta_F = 1$, with $\Delta_S = 16$, $L_S = 32$ and $H = 8$. In all experiments, the slow branch processes time-domain input frames through one \xx{fc} layer and four \xx{gru} layers, each with 64 neurons, followed by another \xx{fc} layer to generate $\epsilon^S$.

Networks are trained for 230 epochs with a batch size of 16. During the first 200 epochs, only the $\mathcal{L}_{\rm SpecMSE}$ loss is used, with a \xx{lr}  starting at 1e-3 and decreasing by 10\% if the loss does not drop for two consecutive epochs. Starting from the 201st epoch, the full loss function (Eq.~\ref{eq:loss}) is applied with weighting factors $[\lambda_1, \lambda_2, \lambda_3, \lambda_4] = [10, 0.5, 1, 5]$. At this point, the \xx{lr} is reset to 1e-4, dropping by 25\% if the loss does not decrease for one epoch. Each network is trained using five different random seeds.

% \subsection{Evaluation Metrics}
Enhancement performance of each network is assessed using six objective measures: 1) PESQ-NB~\cite{rix2001perceptual} and 2) PESQ-WB~\cite{rix2001perceptual} for perceptual quality, 3) STOI~\cite{taal2011algorithm}, 4) ESTOI~\cite{jensen2016algorithm} and 5) HASPI~\cite{kates2014hearing} for intelligibility, and 6) \xx{sisnr} for \xx{snr} improvement, respectively. For all evaluation metrics, higher values denote better \xx{se} performance.

\section{Results}

Table~\ref{tab:2ms} presents \xx{se} results across six evaluation measures for all networks tested under a 2 ms algorithmic latency requirement. Experiments with IDs 0 and 7 serve as baselines, representing the conventional single-branch \xx{se} framework where the entire network must be executed for every input frame. The baseline networks have the same structure as the slow branch of the \method networks. We varied the number of GRU neurons in the baseline networks to adjust their computation cost. Experiments with IDs 1.x to 6.x demonstrate the enhancement performance with the \method framework, using the slow-fast integration methods introduced in Sec.~\ref{sec:slow-fast}. A, B, and C correspond to the \xx{ec}, \xx{film}, and \xx{ssmm} methods, respectively. For reference, we included results of three compute-efficient SE networks from literature (IDs 8, 9, and 10), highlighting that both our baseline and \method networks offer competitive enhancement performance but at a lower computation cost.

Comparing results between ID 0 and ID 3.C, \method - \xx{ssmm} achieves the same enhancement performance as the conventional single-branch framework, but reduces computation cost from 126 M \xxx{mac} to 39 M \xxx{mac} per second, a reduction of approximately 70\%. Comparing ID 7 with ID 6.C, the single-branch framework requires three times more \xxx{mac} to match the performance of the  \method - \xx{ssmm}. These results demonstrate the efficiency of the proposed \method - \xx{ssmm} in significantly reducing computation cost.

Comparing results within x.As, x.Bs, or x.Cs, the \xx{ec} method shows the highest sensitivity to reuse factor $\delta$, with PESQ-WB score dropping by 0.38 (2.45 to 2.07) as $\delta$ increases from 1 to 2. In contrast, the \xx{ssmm} method demonstrates superior robustness, consistently outperforming \xx{ec}- and \xx{film}-based methods across all measures. \xx{ssmm} shows only minor performance decreases for $\delta$ up to 3, maintaining enhancement performance despite less frequent slow branch updates. The slight degradation may be attributed to the increased differences between the speech and noise characteristics modeled by the slow and those of the fast branch inputs as $\delta$ increases. 

Table~\ref{tab:sample-level} presents \xx{se} results under a single sample-level latency constraint. Unlike conventional single-branch methods that require processing the entire network for each incoming input sample, our approach necessitates computing the slow branch only once every 1 ms. This results in an approximate 16-fold reduction in overall computation complexity. Moreover, for our method, if the slow branch is deployed on a remote device, only the fast branch, which consists of 16 parameters, needs to be computed on the local edge device within the 62.5 $\mu$s latency window to realize a real-time online processing system.

\section{Conclusion}

This paper introduces the \method- \xx{ssmm} for efficient low-latency SE. The \method framework combines a heavy, low-frame-rate slow branch for acoustic environment analysis with a lightweight fast branch for enhancing high-frame-rate inputs in the time domain. We propose an \xx{ssmm}-based fast branch that can efficiently utilize acoustic characteristics for enhancement, using only two \xx{fc} layers. Experiments on the VBD dataset demonstrate that \method with \xx{ssmm} significantly reduces computation costs for \xx{se} systems under a 2 ms algorithmic latency constraint. Notably, we present the first fully neural network-based single sample-level latency \xx{se} system, showcasing the \method- \xx{ssmm}'s capabilities. We will apply, in future, \method to those audio signal processing tasks that require even stricter algorithmic latency specifications, such as \xx{anc}.

% References should be produced using the bibtex program from suitable
% BiBTeX files (here: strings, refs, manuals). The IEEEbib.bst bibliography
% style file from IEEE produces unsorted bibliography list.
% -------------------------------------------------------------------------
\bibliographystyle{IEEEbib}
\bibliography{refs}

\begin{thebibliography}{10}

\bibitem{stone2008tolerable}
Michael~A Stone, Brian~CJ Moore, Katrin Meisenbacher, and Ralph~P Derleth,
\newblock ``Tolerable hearing aid delays. v. estimation of limits for open canal fittings,''
\newblock {\em Ear and hearing}, vol. 29, no. 4, pp. 601--617, 2008.

\bibitem{gupta2022augmented}
Rishabh Gupta, Jianjun He, Rishabh Ranjan, Woon-Seng Gan, Florian Klein, Christian Schneiderwind, Annika Neidhardt, Karlheinz Brandenburg, and Vesa V{\"a}lim{\"a}ki,
\newblock ``Augmented/mixed reality audio for hearables: Sensing, control, and rendering,''
\newblock {\em IEEE Signal Processing Magazine}, vol. 39, no. 3, pp. 63--89, 2022.

\bibitem{nicolson2020masked}
Aaron Nicolson and Kuldip~K Paliwal,
\newblock ``Masked multi-head self-attention for causal speech enhancement,''
\newblock {\em Speech Communication}, vol. 125, pp. 80--96, 2020.

\bibitem{pandey2021dense}
Ashutosh Pandey and DeLiang Wang,
\newblock ``Dense {CNN} with self-attention for time-domain speech enhancement,''
\newblock {\em IEEE/ACM Transactions on Audio, Speech, and Language Processing}, vol. 29, pp. 1270--1279, 2021.

\bibitem{fan2022real}
Junyi Fan, Jibin Yang, Xiongwei Zhang, and Yao Yao,
\newblock ``Real-time single-channel speech enhancement based on causal attention mechanism,''
\newblock {\em Applied Acoustics}, vol. 201, pp. 109084, 2022.

\bibitem{oostermeijer2021lightweight}
Koen Oostermeijer, Qing Wang, and Jun Du,
\newblock ``Lightweight causal transformer with local self-attention for real-time speech enhancement.,''
\newblock in {\em Interspeech}, 2021, pp. 2831--2835.

\bibitem{takeuchi2020real}
Daiki Takeuchi, Kohei Yatabe, Yuma Koizumi, Yasuhiro Oikawa, and Noboru Harada,
\newblock ``Real-time speech enhancement using equilibriated rnn,''
\newblock in {\em ICASSP 2020-2020 IEEE International Conference on Acoustics, Speech and Signal Processing (ICASSP)}. IEEE, 2020, pp. 851--855.

\bibitem{cheng2024dynamic}
Longbiao Cheng, Ashutosh Pandey, Buye Xu, Tobi Delbruck, and Shih-Chii Liu,
\newblock ``Dynamic gated recurrent neural network for compute-efficient speech enhancement,''
\newblock in {\em Interspeech}, 2024, pp. 677--681.

\bibitem{luo2018tasnet}
Yi~Luo and Nima Mesgarani,
\newblock ``Tasnet: time-domain audio separation network for real-time, single-channel speech separation,''
\newblock in {\em 2018 IEEE International Conference on Acoustics, Speech and Signal Processing (ICASSP)}. IEEE, 2018, pp. 696--700.

\bibitem{luo2020dual}
Yi~Luo, Zhuo Chen, and Takuya Yoshioka,
\newblock ``Dual-path {RNN: E}fficient long sequence modeling for time-domain single-channel speech separation,''
\newblock in {\em ICASSP 2020-2020 IEEE International Conference on Acoustics, Speech and Signal Processing (ICASSP)}. IEEE, 2020, pp. 46--50.

\bibitem{tu2021two}
Zehai Tu, Jisi Zhang, Ning Ma, Jon Barker, et~al.,
\newblock ``A two-stage end-to-end system for speech-in-noise hearing aid processing,''
\newblock {\em Proc. Clarity}, pp. 3--5, 2021.

\bibitem{gajecki2021binaural}
Tom Gajecki and Waldo Nogueira,
\newblock ``Binaural speech enhancement based on deep attention layers,''
\newblock {\em Proc. Clarity}, pp. 6--8, 2021.

\bibitem{wang2021deep}
Shanshan Wang, Gaurav Naithani, Archontis Politis, and Tuomas Virtanen,
\newblock ``Deep neural network based low-latency speech separation with asymmetric analysis-synthesis window pair,''
\newblock in {\em 2021 29th European Signal Processing Conference (EUSIPCO)}. IEEE, 2021, pp. 301--305.

\bibitem{wang2022stft}
Zhong-Qiu Wang, Gordon Wichern, Shinji Watanabe, and Jonathan Le~Roux,
\newblock ``{STFT}-domain neural speech enhancement with very low algorithmic latency,''
\newblock {\em IEEE/ACM Transactions on Audio, Speech, and Language Processing}, vol. 31, pp. 397--410, 2022.

\bibitem{pandey2023simple}
Ashutosh Pandey, Ke~Tan, and Buye Xu,
\newblock ``A simple rnn model for lightweight, low-compute and low-latency multichannel speech enhancement in the time domain,''
\newblock in {\em Interspeech}, 2023.

\bibitem{mauler2007low}
Dirk Mauler and Rainer Martin,
\newblock ``A low delay, variable resolution, perfect reconstruction spectral analysis-synthesis system for speech enhancement,''
\newblock in {\em 2007 15th European Signal Processing Conference}, 2007, pp. 222--226.

\bibitem{wood2019unsupervised}
Sean~UN Wood and Jean Rouat,
\newblock ``Unsupervised low latency speech enhancement with rt-gcc-nmf,''
\newblock {\em IEEE Journal of Selected Topics in Signal Processing}, vol. 13, no. 2, pp. 332--346, 2019.

\bibitem{srinivas2024knowledge}
Vidya Srinivas, Malek Itani, Tuochao Chen, Emre~Sefik Eskimez, Takuya Yoshioka, and Shyamnath Gollakota,
\newblock ``Knowledge boosting during low-latency inference,''
\newblock in {\em Interspeech}, 2024.

\bibitem{gu2021efficiently}
Albert Gu, Karan Goel, and Christopher R{\'e},
\newblock ``Efficiently modeling long sequences with structured state spaces,''
\newblock {\em arXiv preprint arXiv:2111.00396}, 2021.

\bibitem{orvieto2023resurrecting}
Antonio Orvieto, Samuel~L Smith, Albert Gu, Anushan Fernando, Caglar Gulcehre, Razvan Pascanu, and Soham De,
\newblock ``Resurrecting recurrent neural networks for long sequences,''
\newblock in {\em International Conference on Machine Learning}. PMLR, 2023, pp. 26670--26698.

\bibitem{perez2018film}
Ethan Perez, Florian Strub, Harm De~Vries, Vincent Dumoulin, and Aaron Courville,
\newblock ``Film: Visual reasoning with a general conditioning layer,''
\newblock in {\em Proceedings of the AAAI conference on artificial intelligence}, 2018, vol.~32.

\bibitem{cheng2021fscnet}
Longbiao Cheng, Junfeng Li, and Yonghong Yan,
\newblock ``{FSCN}et: Feature-specific convolution neural network for real-time speech enhancement,''
\newblock {\em IEEE Signal Processing Letters}, vol. 28, pp. 1958--1962, 2021.

\bibitem{martin2018deep}
Juan~Manuel Martin-Donas, Angel~Manuel Gomez, Jose~A Gonzalez, and Antonio~M Peinado,
\newblock ``A deep learning loss function based on the perceptual evaluation of the speech quality,''
\newblock {\em IEEE Signal Processing lLetters}, vol. 25, no. 11, pp. 1680--1684, 2018.

\bibitem{zhang2018training}
Hui Zhang, Xueliang Zhang, and Guanglai Gao,
\newblock ``Training supervised speech separation system to improve stoi and pesq directly,''
\newblock in {\em 2018 IEEE International Conference on Acoustics, Speech and Signal Processing (ICASSP)}. IEEE, 2018, pp. 5374--5378.

\bibitem{valin2018hybrid}
Jean-Marc Valin,
\newblock ``A hybrid dsp/deep learning approach to real-time full-band speech enhancement,''
\newblock in {\em 2018 IEEE 20th international workshop on multimedia signal processing (MMSP)}. IEEE, 2018, pp. 1--5.

\bibitem{schroter2022low}
Hendrik Schr{\"o}ter, Tobias Rosenkranz, Alberto-N Escalante-B, and Andreas Maier,
\newblock ``Low latency speech enhancement for hearing aids using deep filtering,''
\newblock {\em IEEE/ACM Transactions on Audio, Speech, and Language Processing}, vol. 30, pp. 2716--2728, 2022.

\bibitem{shetu2024ultra}
Shrishti~Saha Shetu, Soumitro Chakrabarty, Oliver Thiergart, and Edwin Mabande,
\newblock ``Ultra low complexity deep learning based noise suppression,''
\newblock in {\em ICASSP 2024-2024 IEEE International Conference on Acoustics, Speech and Signal Processing (ICASSP)}. IEEE, 2024, pp. 466--470.

\bibitem{valentini2017noisy}
Cassia Valentini-Botinhao et~al.,
\newblock ``Noisy speech database for training speech enhancement algorithms and tts models,''
\newblock 2017.

\bibitem{veaux2013voice}
Christophe Veaux, Junichi Yamagishi, and Simon King,
\newblock ``The voice bank corpus: Design, collection and data analysis of a large regional accent speech database,''
\newblock in {\em 2013 international conference oriental COCOSDA held jointly with 2013 conference on Asian spoken language research and evaluation (O-COCOSDA/CASLRE)}, 2013, pp. 1--4.

\bibitem{thiemann2013diverse}
Joachim Thiemann, Nobutaka Ito, and Emmanuel Vincent,
\newblock ``The diverse environments multi-channel acoustic noise database (demand): A database of multichannel environmental noise recordings,''
\newblock in {\em Proceedings of Meetings on Acoustics (ICA2013)}. Acoustical Society of America, 2013, vol.~19, p. 035081.

\bibitem{rix2001perceptual}
Antony~W Rix, John~G Beerends, Michael~P Hollier, and Andries~P Hekstra,
\newblock ``Perceptual evaluation of speech quality (pesq)-a new method for speech quality assessment of telephone networks and codecs,''
\newblock in {\em 2001 IEEE International Conference on Acoustics, Speech, and Signal Processing}, 2001, vol.~2, pp. 749--752.

\bibitem{taal2011algorithm}
Cees~H Taal, Richard~C Hendriks, Richard Heusdens, and Jesper Jensen,
\newblock ``An algorithm for intelligibility prediction of time--frequency weighted noisy speech,''
\newblock {\em IEEE Transactions on Audio, Speech, and Language Processing}, vol. 19, no. 7, pp. 2125--2136, 2011.

\bibitem{jensen2016algorithm}
Jesper Jensen and Cees~H Taal,
\newblock ``An algorithm for predicting the intelligibility of speech masked by modulated noise maskers,''
\newblock {\em IEEE/ACM Transactions on Audio, Speech, and Language Processing}, vol. 24, no. 11, pp. 2009--2022, 2016.

\bibitem{kates2014hearing}
James~M Kates and Kathryn~H Arehart,
\newblock ``The hearing-aid speech perception index (haspi),''
\newblock {\em Speech Communication}, vol. 65, pp. 75--93, 2014.

\end{thebibliography}

\end{document}